\documentstyle[aps,prl,floats,epsf,epsfig,psfig]{revtex}

\newcommand{\moreapprox}{\mbox{$>\approx$}}

\newcommand{\anueR}{\mbox{$\overline{\nu}_{eR}$}}
\newcommand{\nubar}{\mbox{$\overline{\nu}_{e}$}}

\newcommand{\anureaction}{\mbox{$\overline{\nu}_e+p\to n+e^{+}$}}

\tighten
\draft
\begin{document}

\twocolumn[\hsize\textwidth\columnwidth\hsize\csname
@twocolumnfalse\endcsname
\title{ Bounds on the Solar Antineutrino total Flux and 
 Energy spectrum from the SK experiment }

\author{E. Torrente-Lujan.}
\address{
IFIC-Dpto. Fisica Teorica. CSIC-Universitat de Valencia. 
Dr. Moliner 50, E-46100, Burjassot, Valencia, Spain.\\
e-mail: e.torrente@cern.ch
}

\maketitle

\begin{abstract}
A search for inverse beta decay electron antineutrinos
has been carried out using the 825 days sample of solar data 
obtained at SK.
The absence of a significant signal, that is, contributions to
the  total SK background and their 
angular variations has set 
upper bounds on  a) the absolute flux of solar antineutrinos
originated from ${}^8 B$ neutrinos 
$\Phi_{\overline{\nu}}({}^8 B)=< 1.8\times 10^5\   cm^{-2}\ s^{-1}$
which is equivalent to an averaged conversion probability bound
of $P<3.5\%$ (SSM-BP98 model)
and b) their differential energy spectrum, the conversion 
probability is smaller than $8\%$ for all $E_{e,vis}>6.5$ MeV going
down the $5\%$ level above $E_{e,vis}\approx 10$ MeV.
It is shown that  an antineutrino flux would have the
net effect of enhancing the SK signal at {\em hep} neutrino
energies. The magnitude of this enhancement would highly
depend on the, otherwise rather uncertain at this moment, steepness  of the
solar neutrino spectrum at these energies. 
\end{abstract}


\vspace{0.5cm}]
\narrowtext

{\bf Introduction.}
 The combined action of  spin flavor precession in a magnetic field
 and  ordinary 
neutrino matter oscillations can produce an observable 
flux of $\anueR$'s from the Sun 
in the case of the neutrino being  a Majorana particle.
In the simplest model, where a thin layer of highly chaotic of 
magnetic field is assumed at the bottom of the convective zone ($R\sim 0.7 R_\odot$), the
antineutrino appearance  probability at the exit of the 
layer can be written as \cite{tor20} (see also Refs.\cite{tor3,tor2,tor5}):
\begin{equation}
P(\nu_{eL}\to  \tilde{\nu}_{eR}  )_f  = \xi
 P(\nu_{eL}\to \nu_{\mu L} )_i,
\end{equation}
where the parameter
$1-2\xi\sim\exp(-4 \Omega^2 \Delta r)$ includes the layer
width( $\Delta r\sim 0.1 R_\odot$) and the r.m.s strength of the
field.
The antineutrino flux could be large if, i.e., the 
 neutrino have passed through a MSW resonance  before arriving to 
the layer. The MSW resonance 
 converts practically all the initial $\overline{\nu}_e$ flux 
into $\overline{\nu}_{\mu}$. The  field finally 
converts them into 
$\overline{\nu}_e$. A fraction  of the $\overline{\nu}_e$
 will be reconverted into  $\overline{\nu}_{\mu}$ by mass 
oscillations but this reconversion
is limited in this case by the chaotic character of the process.

 Water Cerenkov detectors such as Kamiokande and
 SuperKamiokande  (SK) which are 
sensitive to the $\nu_e-e$ elastic interaction are 
also capable of detecting these 
$\overline{\nu}_e$'s coming from the sun. Forthcoming solar 
neutrino experiments, such 
as SNO and Borexino are also expected to have  a  high 
sensitivity to them.
The specific signature of electron antineutrinos in proton
 containing materials is the inverse beta decay process:
 $\overline{\nu}_e+p\to n+e^{+}$, which produces 
almost isotropical   monoenergetic positrons with a relatively high 
cross section. Antineutrino events would contribute  to the 
 background to forward-peaked neutrino solar events.

The residual 
angular correlation  between the direction of the incident neutrino 
 and the resulting electron  was proposed already in Ref.\cite{reines} as a way of detecting solar neutrinos.
More recently, it was again 
suggested \cite{fio1} 
the same procedure to separate statistically
 antineutrino events in the SK sample.
In practice however, as it was pointed out in Ref.\cite{vogel} and 
we show in detail in this work, the angular distribution 
is of a more complicated nature 
than was naively  assumed in Ref.\cite{fio1}.
Nevertheless, as we show in the detailed analysis presented here, meaningful
bounds on solar antineutrino fluxes and appearance probability can be 
obtained from the SK data. 

Solar antineutrinos could have been already detected at SK: due
to the anomalous, forward-peaked at the higher
energies, angle distribution the 
excess of high energy solar neutrino events observed in SK above
13 MeV, the
{\em hep} spectrum anomaly \cite{sk1,bah2}, could be
 explained, albeit partially, as an excess of positrons 
coming from inverse beta decay appearing  with a direction
which fakes that one of real solar neutrino electrons.

{\bf Antineutrino Cross sections.}
In the limit of infinite nucleon mass, the 
cross section  for the reaction 
\anureaction\  is given by
\cite{zacek,reines} $\sigma(E_{\overline{\nu}})=c_1 E_e p_e$.
The transition matrix element 
$c_1=2\pi^2 \log 2/ m_e^5 ft_{1/2}$
is  expressed in terms of the free neutron
decay $ft_{1/2}$ value, where the phase-space factor
$f=1.71465\pm 0.00015$ follows from calculation \cite{zacek21}
and
$t_{1/2}=614.6\pm 1.3$ sec is latest published value for 
the neutron half-life \cite{PDG99}.
With $M_n$, $M_p$ and $m_e$ being the masses of neutron, proton and
electron, respectively 
and $\Delta M=M_n-M_p\simeq 1.293 $ MeV.

The energy $E_{\overline{\nu}}$ of the 
incident neutrino is related to that one of the positron 
$E_{e^+}$ by:
$E_{\overline{\nu}}^0=E_{e^+}+\Delta M$.
From the values above, we obtain: $c_1=(9.54\pm 0.02)\times 10^{-44}$ 
cm$^2$/MeV$^{2}$.
Corrections to  the total cross-section due to weak magnetism
 arising from the difference in the anomalous magnetic moments of the
neutron and proton and radiative corrections including internal
bremsstrahlung (see Ref.\cite{zacek} and references therein)
 are very small and can be neglected for our purposes.

Direct recoil corrections to the total cross section can also be neglected in principle.
However, recoil corrections are potentially important 
in relating the positron and antineutrino energies in order to 
evaluate the antineutrino flux. 
The  positron spectrum  is not monoenergetic in this case 
and one has to integrate over the positron angular 
distribution to obtain the positron yield.
At the SK experiment,because the antineutrino 
flux $\Phi(E_\nu)$ would typically decrease quite 
rapidly with energy, the lack of adequate corrections 
will systematically overestimate the   positron yield.   
For the SK case and taking into account the SSM-BP98 ${}^8 B$ spectrum, 
the effect  decrease the positron 
yield by 2-8\% at the main visible energy range $\sim 6-10$ MeV.
The positron yield could decrease up 50\% at {\em hep} 
neutrino energies, a region where incertitudes in the total 
and differential spectrum are of comparable size or larger.
 Finite energy resolution smearing will however diminish  this correction 
when integrating over large enough energy bins: in the range $6.5-20$ MeV
 the net positron suppression is estimated to be at the $5\%$ level, increasing up $20\%$ at {\em hep} energies.

Indirect recoil effects through the antineutrino flux spectrum are also 
present in the angular distribution as we discuss below. 

{\bf The angular distribution.}
For low antineutrino energies, the positron angular distribution 
is well-described by the expression
\begin{eqnarray}
\frac{d\sigma}{d \cos\theta}=\sigma(E_\nu)\frac{1}{2}\left (\vphantom{a^{(0)}} 
1-v_e \alpha(E_\nu)  \cos\theta\right )
\label{e7004}
\end{eqnarray}
where $\theta$ is the lab angle between the antineutrino and positron 
 directions, $v_e$ is the positron velocity.
Except  near threshold, this velocity is nearly unity and can be
ignored.
For a linear distribution
 as that given by Eq.(\ref{e7004}), the values for the average 
cosine and the $\alpha$ coefficient are related by: 
$\alpha=3\langle \cos\theta\rangle/v_e$.
In the limit of infinite  nucleon mass, the  coefficient 
$\alpha=\alpha^{(0)}\equiv
(\eta^2-1)/(3 \eta^2+1)\approx 0.10$
is independent of $E_\nu$, 
  $\eta\equiv g_A/g_V$ is the ratio between  axial and vector couplings of the
neutron. Thus the angular distribution of the positrons is weakly 
backwards, independent of the energy above the threshold region.

\begin{figure}[h]
\centering
\epsfig{file=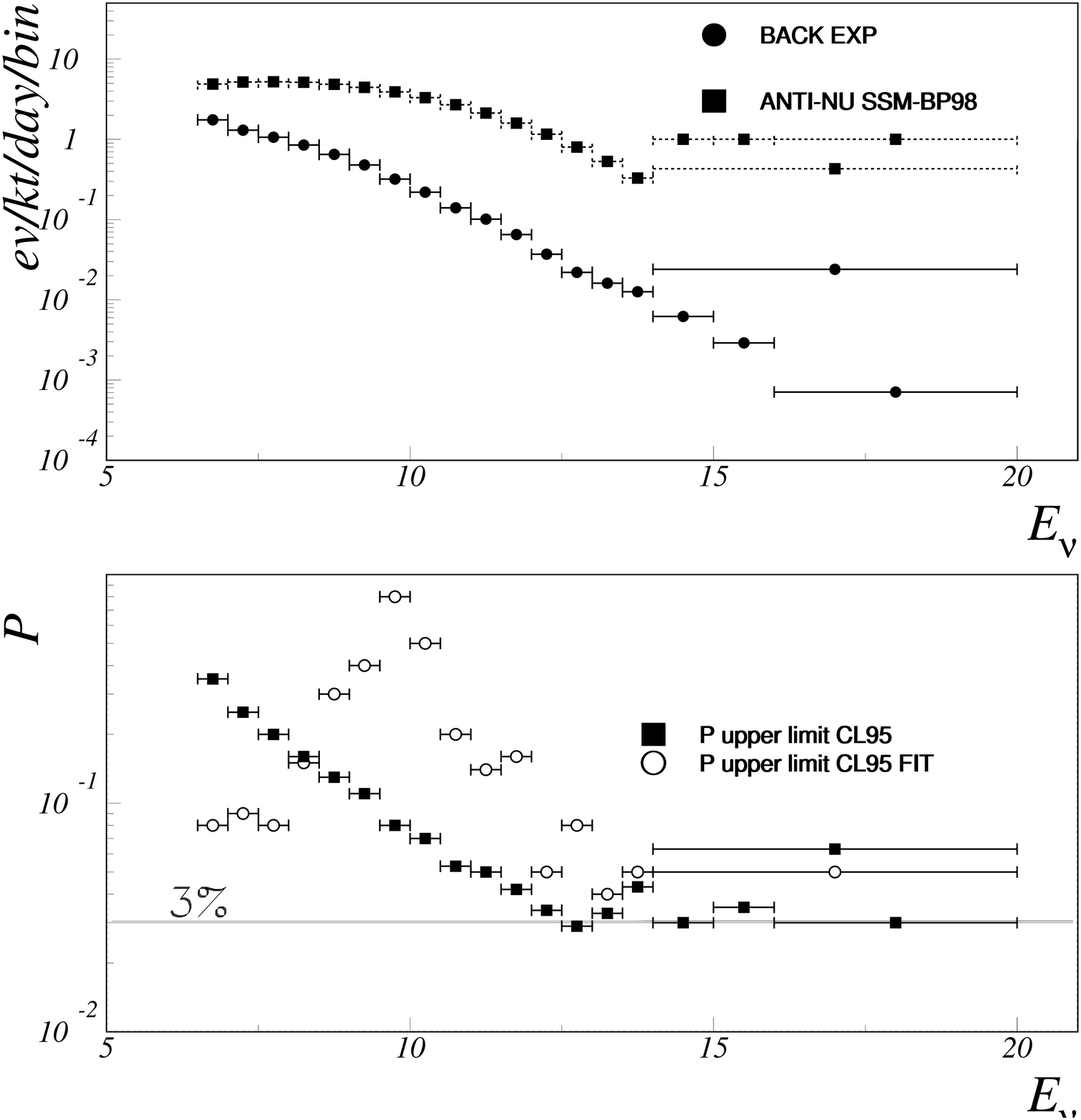,height=7cm,width=7cm}
\caption{Top: observed flat background and maximal
expected positron signal. Bottom: upper limits on antineutrino
conversion probability obtained from global counting (solid circles) and
angle fit (empty circles).}
\label{f2}
\end{figure}

The situation greatly changes when 
weak magnetism and $O(1/M)$ recoil corrections are kept as is shown 
in Ref.\cite{vogel}. At higher energies, terms proportional to higher 
powers of $\cos\theta$ appear.
The average cosine is changed by a noticeable
 amount.  It changes sign for not-so-high 
energies ($E_\nu\sim 12-13$ MeV), the distribution
 becomes then forward peaked.
At first order $O(1/M)$, we have  
($E_\nu\moreapprox 5$ MeV)
\begin{eqnarray}
\frac{d\sigma^{(1)}}{d \cos\theta}=\frac{\sigma(E)}{2}\left ( 
1-\alpha^{(1)} \cos\theta+\beta \cos^ 2\theta\right )
\label{e3006}
\end{eqnarray}
where the coefficients $\alpha^{(1)},\beta$ can be read off 
from Eq.(19) in Ref.\cite{vogel}. The average cosine can be computed from here yielding the following 
expression which is  an accurate approximation valid  from
 the threshold up $E_\nu\approx 100$ MeV
($f_2=\mu_p-\mu_n=3.706$ is 
the anomalous nucleon magnetic moment,$\eta'=4 \eta/3(1-\eta))$ \cite{vogel}:
\begin{eqnarray}
\langle \cos\theta\rangle^{(1)}&=&  v_e \alpha^0/3
+\left ( 1+ (1+f_2)\eta' a^0\right ) E_\nu/M\nonumber\\
&\simeq& -0.034\  v_e +2.55\times 10^{-3} E_\nu (MeV). 
\label{e5678}
\end{eqnarray}
The angular differential cross section (\ref{e3006}) is shown in 
 Fig.(\ref{f2}) for different neutrino energies.
It is important to notice that   the angular distribution is still 
linear to a high degree. 
We can rewrite an effective  expression which 
is a very good approximation 
to the first order expression in all the energy range.
It can be shown, by 
obvious arguments or, if prefered, by a rigorous Least Square approach, 
that the best linear fit to the quadratic expression (\ref{e3006}) is given
by
\begin{equation}
\frac{d\sigma^{(1,lin)}}{d \cos\theta}=\frac{\sigma(E)}{2}\left ( 
1-\alpha^{(1)}_{eff}(E) \cos\theta\right ).
\label{e8745}
\end{equation}
where the effective constant is:
\begin{equation}
\alpha^{(1)}_{eff}(E_\nu)= 3 \langle\cos\theta\rangle^{(1)}
\label{e4523}
\end{equation}
and $\langle\cos\theta\rangle^{(1)}$ is given by expression (\ref{e5678}).
The angular distribution can be completely  parametrized by a single quantity which is 
a linear function of the neutrino energy.
\begin{figure}[h]
\centering\hspace{0.8cm}
\epsfig{file=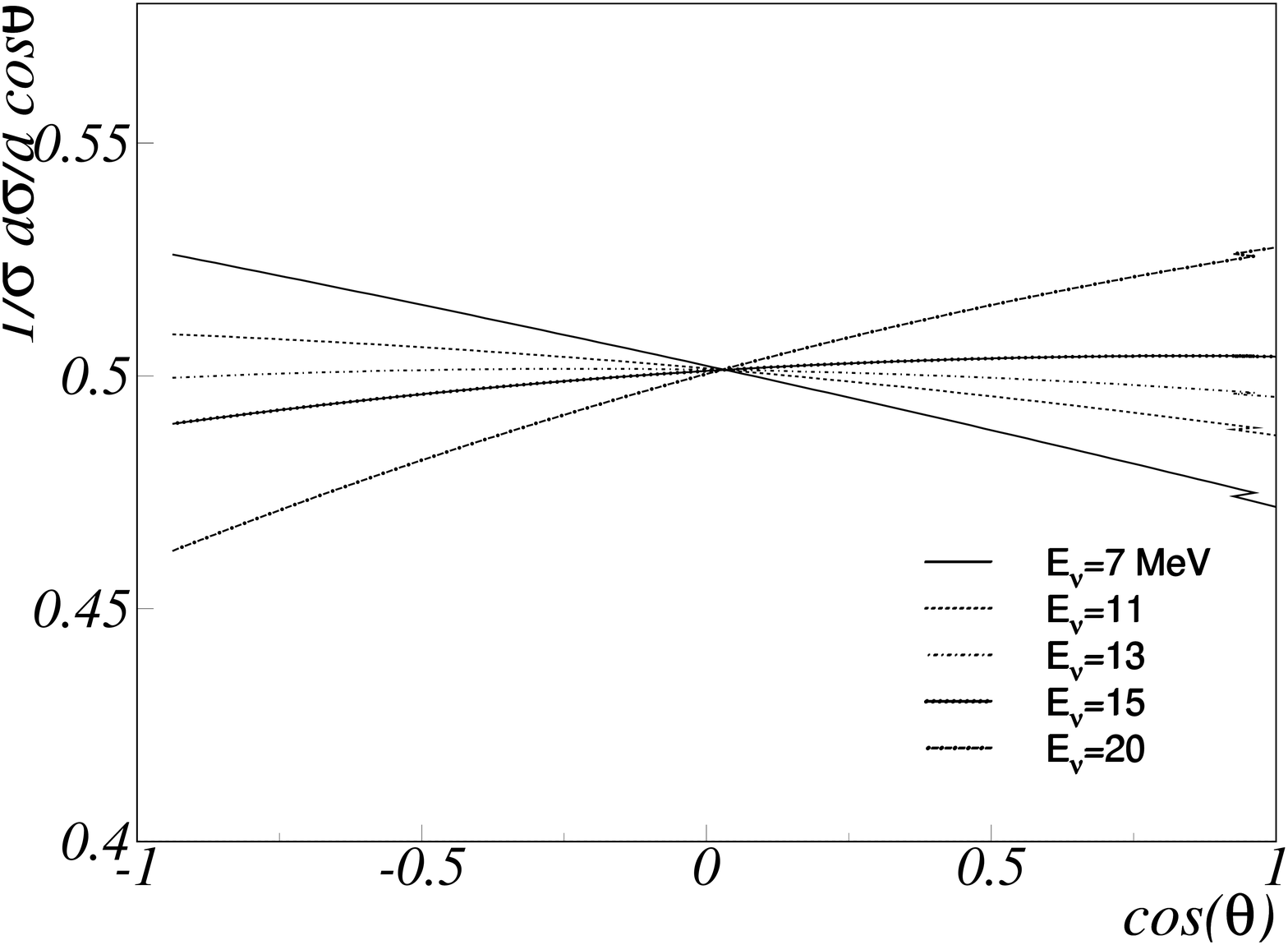,height=4cm,width=6cm}
\caption{The angular distribution, Eq.(\protect\ref{e3006}),
for different antineutrino energies. Above $E_\nu\sim 13 $ MeV the 
distribution is peaked forwards.}
\label{f1}
\end{figure}

Recoil corrections to the angle positron distribution are
important when relating positron and antineutrino energies.
In fact they become the dominant effect at the highest SK 
energies.
The observable quantity is the positron angular distribution
 as a function of the  positron energy itself 
$dN_e/d\cos\theta(E_e)\sim d\sigma/d\cos\theta(E_e)\times \Phi_\nu[E_\nu(cos\theta)]$. 
with $E_\nu(cos\theta)=E_\nu^0/(1-E_{e^+}(1-\cos\theta)/M)$.
At the same positron energy, the difference of the energy
of the parent antineutrino for positrons respectively 
emitted at backward and forward energies is of order $O(1/M)$ but increasing quadratically with the positron energy:
\begin{eqnarray}
\Delta E_{\overline{\nu}}\simeq -2 E_e (E_e+\Delta M)/M,
\end{eqnarray}
at $E_e\approx 12-15$ MeV, the difference 
$\Delta E_{\overline{\nu}}
\approx 0.50 $ MeV.
The signal difference seen between the forward and backward
directions 
$A\equiv\frac{F-B}{F+B}$
is related to the average cosine. 
For our angle distribution, we have
$A=(3/2) \langle\cos\theta\rangle=\alpha/2$.
We have, schematically,
\begin{eqnarray}
\alpha_{N}\propto\Delta N_{e^+}\sim \int dE_e 
\mid \frac{\partial \Phi}{\partial E_\nu}\mid
\Delta E_{\overline{\nu}}.
\end{eqnarray}
It is estimated that, for 1 MeV positron energy bins at $E_e\sim 13$ MeV, the effective angular positron distribution is
forward peaked with $\alpha_{N}\sim 0.4-0.5$ (to be compared with
 values in Fig.\ref{f1}). 

{\bf Antineutrino asymmetry and high energy 
 neutrino enhancement.}
 In the SuperKamiokande experiment,
the extraction of the solar neutrino signals from the final sample is
basically carried out \cite{hirothesis} 
using directional correlation of the 
events  to the direction of the sun, since the 
recoil electrons keep 
the directionality of the incoming neutrino.
The distribution of the angle between the 
reconstructed direction and the direction of the sun
presents a peak in the sun direction, $\cos\theta=1$. 
The flat component
for $\cos\theta<\approx 0.4$ is considered as background. 
the excess above this baseline is  defined as the solar
 neutrino signal.

In order to count number of the solar neutrino events, 
a maximum likelihood method is employed. 
Signal and 
background  probability functions ($P_s,P_{bg}$) are defined which 
depend on angles and energies,
due to the finite, energy dependent, angular resolution 
of the detector.
The background probability density is obtained from the 
data. In the ideal case, the function does not depend 
on $\cos\theta$ and $P_{bg}=1/2$.
In the real case,
backgrounds, such as $\gamma$-rays, which may have strong
directionality in the detector coordinate system are 
eliminated by fitting a multi-order, energy dependent,
polynomial to the detector coordinate distribution and 
and mapping it back to the direction of the sun.
Clearly, this kind of methods does  not eliminate 
a systematic forward-peaked background as antineutrinos which is virtually 
un-distinguishable from the real neutrinos.
The net  effect being that the  
real signal is systematically 
underestimated at low energies and, to a much higher degree, overestimated at higher visible energies. 
As it is implied by the results to be presented below where
we use the standard model  neutrino spectrum,
the net effect is too small in practice to explain the 
experimental excess \cite{sk1,bah2}.
It is intriguing anyway  that the vanishing  
and sign change of the 
antineutrino asymmetry occurs precisely at right position of the
high energy neutrino spectrum $E_\nu\approx 12.5$ MeV.
Irrespective of the total flux, 
the situation could change if the 
spectrum profile, through the quantity 
$\mid\partial \Phi/\partial E_\nu\mid$,
highly deviates from the assumed standard value.

{\bf Experimental angular distributions.}
The direction of the positron from antineutrino capture 
is smeared away by the angular 
resolution of the detector. This angular resolution is energy 
 dependent and due to it,
the effective, experimentally detectable,
 asymmetry parameter $\alpha_{eff}$ is slightly
 smaller than the theoretical one and includes an additional
 small energy dependence.

We have performed a simple Monte Carlo simulation in order to 
compute the influence of such effect in the SK data.
We have found convenient to  parameterize  the 
SK angular resolution in terms of the Beta distribution
$\epsilon (\theta)_{m,n}\sim \theta^{m-1} (1-\theta)^{n-1}$,
the parameters $m,n$ has been obtained from a fit to the 
data presented in Ref.\cite{hirothesis,nakahata}. 
In particular the variance corresponding to the Beta 
distribution 
can directly be obtained from a fit to the energy dependent
angular resolution presented in  Ref.\cite{hirothesis}.
Full concrete expressions for $\sigma_E$ 
will be presented in Ref.\cite{newtor}.
We have found that   the 
energy dependence of $\alpha_{eff}$ can be parametrized as a simple
linear expression. 

The main conclusion is that 
the net effect of the finite angular resolution is the smearing of the 
initial angular distribution consistent  with a reduction 
of no more  $15\%$ on $\alpha_{eff}$ at the lowest energies observable at SK.
At the highest energies, the effect of the detection angular 
resolution decreases becoming negligible.

{\bf Expected antineutrino flux.}
The average number of positrons $N_i$ which are detected per
visible energy bin $\Delta E_i$ is given by the convolution of 
different quantities
\begin{eqnarray}
&&N_i= Q_0 \int_{\Delta E_i}dE_e \int_0^\infty dE_e^r \epsilon(E_e)f(E_e,E_e^r)
\nonumber\\ 
& &\times \int_{E_e^r}^\infty dE_{\overline{\nu}} \overline{F}(E_{\overline{\nu}})
 \sigma (E_{\overline{\nu}},E_e^r)
\label{e3466} 
\end{eqnarray}
where $Q_0$ is a normalization constant  accounting for the fiducial volume
 and live time,  
$\overline{F}(E_{\overline{\nu}})$ is the flux of solar antineutrinos
 per unit energy at 
the detector.
When ignoring recoil effects
the antineutrino capture cross section
$\sigma (E_{\overline{\nu}},E_e^r)$ 
is simply given by
$\sigma (E_{\overline{\nu}},E_e^r)=\sigma (E_{\overline{\nu}})\delta(E_{\overline{\nu}}-E_e^r)$ with  
$\sigma (E_{\overline{\nu}})$    given as before.
The functions $\epsilon(E_e)$ and 
$f(E_e,E_e^r)$ are, respectively, the detection
 efficiency and  the energy resolution function which includes
  trigger and VD efficiencies. Expressions for the functions $\epsilon,f$
has been obtained by us using the data presented in
 Refs.\cite{hirothesis,nakahata}. Further analysis and 
explicit expressions  can be found in Ref.\cite{newtor}.
The antineutrino flux, production probability
$P_{\nubar}(E_\nu)$  
and the solar-originated neutrino flux $F$ are related by 
$\overline{F}=F\times  P_{\nubar} (E_\nu)$.

{\bf Results.}
We have analyzed  the 
data corresponding to the full energy range $6.5-20$ MeV 
from the first 504 days of 
operation of SK as reported in  
Ref.\cite{hirothesis} 
as well as 
the preliminary LE-trigger data corresponding to the first 703 and 825 days
of operation \cite{smy,naka2}.
In addition we have analyzed data corresponding to 
individual energy bins of 0.5 MeV
interval 
as presented in  Ref.\cite{hirothesis}. Note that, in the 
high energy end, individual 
energy bins $14-15$, $15-16$ and $16-20$ MeV are also considered.
    
{\em Limits from integrated  background data.}
The results obtained from the analysis of the angle-integrated background data
over the full 
energy range are summarized in Table~(\ref{t1}) (see second column).
From the condition $N_i< Back_{exp}$, a model independent upper limit on the 
flux of antineutrinos originated from ${}^8 B$ neutrinos is 
obtained $\Phi_{\overline{\nu}}({}^8 B)< 7.5\times 10^5\   cm^{-2}\ s^{-1}.$
Note that this limit would include antineutrinos with energies covering 
  all the ${}^8 B$ energy range.
This number is equivalent to  
an upper bound $P< 14$\% CI95\% on the energy averaged
 neutrino-antineutrino conversion probability (SSM-BP98). 
Note that much more strict bounds will be obtained below.

\begin{figure}[h]
\centering
\epsfig{file=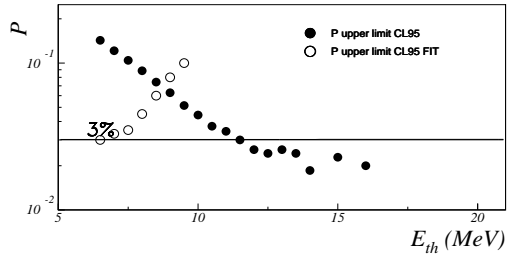,height=7cm,width=7cm}
\vspace{-2cm}
\caption{Upper limits on antineutrino
conversion probability as a function of $E^{vis}_{e^+,th}$. As Fig.(\protect\ref{f2})}
\label{f3}
\end{figure}

The corresponding results obtained from the analysis of individual 
energy bins are shown in the Figs.(\ref{f2}). In Fig.(\ref{f2}) (top)
we show per each positron visible energy bin (width $\pm 0.25 MeV$,
 except higher energies)  the 
 observed flat background, the maximal SSM (BP98, see Ref.\cite{BP95}) expected positron signal
(supposing $P=1$). In the bottom figure (solid circles) we present 
 upper limits on $P$ which are inferred.
We observe that the average conversion probability 
is below $8-10\%$ for energies above 9 MeV and below the $5\%$ line 
for visible energies above 10 MeV.
In Fig.(\ref{f3}) we present the same information in a slightly 
different, smoother way, for integrated 
bins above a certain threshold: the solid circles represent
  upper limits on P obtained from the consideration of the
global background.

The limits from global background are complemented, specially in the 
lower energies, with those obtained from consideration of the 
their angle distribution as follows.

\begin{table}[h]
\centering
\begin{tabular}{|c|c|c|}
Live Time & $Back_{exp}$  & $\Phi_{\overline{\nu}}({}^8 B)    $   \\ \hline
504 days  & $7.05\pm 0.03$  &     $<7.7\times 10^5$       \\
703 days   & $6.98\pm 0.03$  &    $<7.6\times 10^5$          \\
825 days   & $6.87\pm 0.02$  &    $<7.5\times 10^5$     
\end{tabular}
\caption{
SK live time data.
$Back_{Exp}$(evt/kt/day): observed flat background. Limits on the 
antineutrino flux $\Phi_{\overline{\nu}}({}^8 B)$  (cm$^{-2}$ s$^{-1}$).  
  We have supposed for the antineutrino a solar neutrino spectrum
with a  $^8$B absolute flux
$\Phi_\nu({}^8 B)=5.15^{+0.98}_{-0.72}\times 10^6$ cm$^{-2}$ s$^{-1}$
\protect\cite{BP95}.
This absolute flux would yield a maximal quantity of 
48.2 evt/kt/day  antineutrinos observable in SK.}
\label{t1}
\end{table}

{\em Limits from fit to the angular distribution.}
Bounds on the  antineutrino flux  have been extracted  from the 
fit to the counting yield in the angular region where ``background'' events
 from $\nu e$ interactions can be ne\-glec\-ted. 
For this purpose the use of the linear 
expression given by Eq.(\ref{e8745}) instead of the, 
only marginally more exact Eq.(\ref{e3006}), is especially 
advantageous because allows for a trivial averaging over a finite energy
range. The averaging of a complex energy-dependent  expression is traded 
off by the averaging of a single parameter which depends linearly on the 
energy. 
An small solar neutrino contribution is still expected at 
 $\cos\theta\sim 0.5$. 
We have explicitly checked that the results to be presented below, which 
use the available information about the slope sign, 
are independent of the 
concrete angle range used for fitting, in practice we have found 
rather stable results when we fit  any number of bins up cosine values
in the range $\cos\theta<0.25-0.5$. 
\begin{table}[h]
\centering
\begin{tabular}{|c|c|c|c|}
Live Time & $(P \langle \alpha\rangle)_{\exp}$ & $P<P_0$    &$\Phi_{\overline{\nu}}({}^8 B)    $   \\ \hline
504 days& $1.6\times 10^{-3}$& 0.075& $< 3.4 \times 10^5$     \\
700 days& $8.8\times 10^{-4}$& 0.040& $< 2.1 \times 10^5$    \\
825 days& $7.7\times 10^{-4}$& 0.035& $< 1.8 \times 10^5$    
\end{tabular}
\caption{$(P\langle\alpha\rangle)_{\exp}$: slope 
fit to the experimental angle distributions ($E_{th}>6.5$ MeV),
CI95\% (unified Feldman-Cousins  Approach with the boundary condition
$(P\langle\alpha\rangle)_{Exp}>0$ ). 
Limits on the antineutrino probability supposing 
$\langle\alpha_{eff}\rangle=0.022$ (see text).
Bounds on the 
antineutrino flux $\Phi_{\overline{\nu}}({}^8 B)$  (cm$^{-2}$ s$^{-1}$) (see 
Table(\protect\ref{t1})).}
\label{t5}
\end{table}

In order to draw a limit 
corresponding to  a  visible   energy bin of width  $\Delta E_i$,
the fitted  slope $b$ must be compared with the expected value 
$\alpha_{\exp}= Q_0/M\ \langle \alpha_{eff}^{(1)}
\rangle_{\Delta E_i} $,
where
$\langle \alpha_{eff}^{(1)}\rangle_{\Delta E_i}$ is the 
energy average of Eq.(\ref{e4523}) over the 
energy range $\Delta E_i$ and $M$ is the number of bins in which 
the full angle range  is partitioned.
From Eqs.(\ref{e3006}) and (\ref{e4523}), we obtain
$$
<\alpha_{eff}^{(1)}>_{\Delta E_i}= 
-0.092+ 2.55\times 10^{-3} \langle E_{e^+}\rangle_{\Delta E_i}. 
$$
For example, for the full range $6.5-20$ MeV, 
$\langle E_{e^+}\rangle\simeq 10.0$ MeV and the value for the 
observable distribution parameter
$\langle \alpha_{eff}^{(1)}\rangle \simeq -0.022$. 
Note that this value is far from the 
naive value for $\alpha^0=-0.1$ quoted above. Note in addition that 
 detector  angular resolution effects have 
been ignored here, according to estimations
(see MC simulations above)
they would induce a further, modest, smearing out  of about $10-15\%$ which
conservatively has been included in the results.
At higher energies and for small bin widths the recoil
corrections are important and have throughly  been included.

The results obtained from the fit to  the full 
energy range distributions are shown in Table (\ref{t5}).
An important  improvement with respect to the bounds  derived from the
global background is obtained.
The model independent upper limit on the 
${}^8 B$ antineutrino flux is now
$\Phi_{\overline{\nu}}({}^8 B)< 1.8\times 10^5\   cm^{-2}\ s^{-1},$
which corresponds to  an upper bound $P<$ 3.5\% CI95.
In order to extract CI intervals, 
we have used the known information
about the sign of the expected slope: the 
unified Feldman-Cousins  approach for dealing with boundary problems
\protect\cite{feldman}  
has been used for this purpose.

A similar analysis was performed for each of the 
individual energy bins and for the cumulated distributions.
The results are shown in Figs.(\ref{f2})(bottom figure, empty circles) and
(\ref{f3}) (empty circles).
We see from 
the figures the complementary nature of both type of analysis: 
sensible limits are obtained only for energies down 10 MeV, the region where
the bounds obtained from consideration of the global background are 
weaker. 
One draws a combined limit of $P<8\%$ CI95\% for the full energy range. The 
bounds are in fact  lower than the  $5\%$ level 
 for the great part of the interval.

In summary, there are no indications of electron antineutrinos in the
 up-to-date SK data. This negative result
sets upper bounds on the total and 
differential solar antineutrino spectrum.
We obtain an upper bound  $P_{\nu\to\overline{\nu}}<\approx 3.5\%$ CI95\%
above the 6.5 MeV threshold.
As a function of the energy, the stricter limits ($\sim 2\%$)
are obtained for the highest positron
visible  energies.

This work has been supported by DGICYT under Grant 
 PB95-1077 and by  a DGICYT-MEC contract  at Univ. de Valencia.


\end{document}